\documentclass[11pt]{article}
\usepackage{bm} 
\usepackage{color}
\usepackage{graphicx}
\usepackage{amsmath}
\usepackage{authblk}

\newcommand{\argmin}{\mathop{\rm arg~min}\limits}

\makeatletter
\begin{document}
\title{Geometry of Information Integration}
\author[1,2]{Shun-ichi Amari}
\author[3]{Naotsugu Tsuchiya}
\author[1,2]{Masafumi Oizumi}
\affil[1]{RIKEN Brain Science Institute}
\affil[2]{Araya Inc.}
\affil[3]{School of Psychological Sciences, Monash University }
\date{}

\maketitle

\begin{abstract}
Information geometry is used to quantify the amount of information integration within multiple terminals of a causal dynamical system. Integrated information quantifies how much information is lost when a system is split into parts and information transmission between the parts is removed. Multiple measures have been proposed as a measure of integrated information. Here, we analyze four of the previously proposed measures and elucidate their relations from a viewpoint of information geometry. Two of them use dually flat manifolds and the other two use curved manifolds to define a split model. We show that there are hierarchical structures among the measures. We provide explicit expressions of these measures.

\end{abstract}

\section{Introduction}

It is an interesting problem to quantify how much information is integrated in a multi-terminal causal system. The concept of information integration was introduced by Tononi and colleagues in Integrated Information Theory (IIT), which attempts to quantify the levels and contents of consciousness \cite{Tononi2004,Balduzzi2008,Oizumi2014}. Inspired by  Tononi's idea, many variants of integrated information have been proposed \cite{Barrett2010,Tegmark2016,Oizumi2016PLoS,Oizumi2016PNAS}. From a different perspective from IIT, Ay independently derived the same measure as integrated information proposed in \cite{Barrett2010} to quantify complexity in a system \cite{Ay2001,Ay2015}. 

In this paper, we use information geometry \cite{Amari2016} to clarify the nature of various measures of integrated information as well as the relations among them. Consider a joint probability distribution $p({\bm{x}}, {\bm{y}})$ of sender $X$ and receiver $Y$, where ${\bm{x}}$ and ${\bm{y}}$ are vectors consisting of $n$ components, denoting actual values of $X$ and $Y$. Here, ${\bm{y}}$ is stochastically generated depending on ${\bm{x}}$. That is, information is sent from the sender $X$ to the receiver $Y$. We consider a Markov model, where ${\bm{x}}_{t+1}$ (=${\bm{y}}$) is generated from ${\bm{x}}_t$ (=${\bm{x}}$) stochastically by transition probability matrix $p \left({\bm{x}}_{t+1}|{\bm{x}}_t \right)$. In this way, we quantify how much information is integrated within a system through one step of state transition.

To quantify the amount of integrated information, we need to consider a split version of the system in which information transmission between different elements are removed, so that we can compare the original joint probability with the split one. The joint probability distribution of a split model is denoted by $q({\bm{x}}, {\bm{y}})$. We define the amount of information integration by the minimized Kullback-Leibler (KL) divergence between the original distribution $p({\bm{x}}, {\bm{y}})$ and the split distribution $q({\bm{x}}, {\bm{y}})$,
\begin{equation}
 \Phi = \min_q D_{KL} \left[ p({\bm{x}}, {\bm{y}}) : q({\bm{x}}, {\bm{y}})\right], 
\end{equation}
which quantifies to what extent $p({\bm{x}}, {\bm{y}})$ and $q({\bm{x}},{\bm{y}})$ are different. Minimizing KL-divergence means selecting the best approximation of the original distribution $p({\bm{x}}, {\bm{y}})$ among the split distributions $q({\bm{x}}, {\bm{y}})$. 

We need to search for a reasonable split model. For each distinct version of of split models, corresponding measure of integrated information can be derived \cite{Ay2001,Ay2015,Barrett2010,Oizumi2016PLoS,Oizumi2016PNAS}. The present paper studies four reasonable split models and the respective measures of integrated information. Among the four integrated information, $\Phi_G$, the geometric $\Phi$ defined in \cite{Oizumi2016PNAS}, is what we believe the most reasonable measure for information integration in a sense that it purely quantifies causal influences between parts, although the others have their own meanings and useful characteristics.

\section{Markovian Dynamical Systems}

We consider a Markovian dynamical system
\begin{equation}
 {\bm{x}}_{t+1} = T {\bm{x}}_t
\end{equation}
where ${\bm{x}}_t$ is the state of the system at time $t$ and ${\bm{x}}_{t+1}$ is the state at the next time step $t+1$, which are vectors consisting of $n$ elements. $T$ is a state transition operator, which is represented by the conditional probability distribution of the next state ${\bm{x}}_{t+1}$ given the current state ${\bm{x}}_t$, $p({\bm{x}}_{t+1}|{\bm{x}}_t)$. $p({\bm{x}}_{t+1}|{\bm{x}}_t)$ is called a transition probability matrix. Throughout this paper, we will use ${\bm{x}}$ for $\bm{x}_t$ and $\bm{y}$ for $\bm{x}_{t+1}$ for the ease of notation. 

Given the probability distribution of ${\bm{x}}$ at time $t$, $p({\bm{x}})$, the probability distribution of the next state,  $p({\bm{y}})$, is given by
\begin{equation}
 p({\bm{y}}) = \sum_{\bm{x}} p\left({\bm{y}}|{\bm{x}}\right) p({\bm{x}}),
\end{equation}
and the joint probability distribution is given by
\begin{equation}
 p({\bm{x}}, {\bm{y}}) = p({\bm{y}}|{\bm{x}})p({\bm{x}}).
\end{equation}
Throughout the paper, we will use  $p({\bm{x}})$ and $p({\bm{y}})$ to mean $p_X({\bm{x}})$ and $p_Y({\bm{y}})$, which explicitly and accurately denote $X$ and $Y$. 


The state ${\bm{x}}$ is supported by $n$ terminals and information at terminals $x_1, x_2, \cdots, x_n$ are integrated to give information in the next state ${\bm{y}}= \left(y_1, \cdots, y_n \right)$, so that each $y_i$ depends on all of $x_1, \cdots, x_n$. We quantify how much information is integrated among different terminals through state transition. All such information is contained in the form of the joint probability distribution $p({\bm{x}}, {\bm{y}})$. We use a general model $\mathcal{M}$ to represent $p({\bm{x}}, {\bm{y}})$, called a full model, which is a graphical model where all the terminals of sender $X$ and receiver $Y$ are fully connected. We consider the discrete case, in particular the binary case, in which $x_i$ and $y_i$ are binary taking values of 0 or 1, although generalization to other cases (e.g., continuous, more discretization steps than binary) is not difficult. We also study the case where random continuous variables are subject to Gaussian distributions.

In order to quantify the amount of information integration, we consider a ``split model'' $\mathcal{M}_S$, where information transmission from one terminal $x_i$ to the other terminals $y_j \; (j \ne i)$ is removed. Let $q({\bm{x}}, {\bm{y}})$ be the joint probability distribution of ${\bm{x}}$ and ${\bm{y}}$ in a split model. The amount of information integration in $p({\bm{x}}, {\bm{y}})$ is measured by the KL-divergence from $p({\bm{x}}, {\bm{y}})$ to $M_S$, that is, the KL-divergence from $p({\bm{x}}, {\bm{y}})$ to 
$q^{\ast}({\bm{x}}, {\bm{y}})$,  which is a particular instantiation of the split model and is the one that is closest to $p({\bm{x}}, {\bm{y}})$. Integrated information is defined as the minimized KL-divergence between the full model $p$ and the split model $q$ \cite{Oizumi2016PNAS}, 
\begin{align*}
 \Phi &= {\mathop{\min}_{q \in \mathcal{M}_S}} D_{KL} \left[p({\bm{x}}, {\bm{y}}):q({\bm{x}}, {\bm{y}})\right], \\
 &= D_{KL} \left[p({\bm{x}}, {\bm{y}}):q^*({\bm{x}}, {\bm{y}}) \right].
\end{align*}
Depending on various definitions of ``split'' model $\mathcal{M}_S$, different measures of integrated information can be defined. Below, we elucidate the nature of the other three candidate integrated information and their relations.

\section{Stochastic Models of Causal Systems}
\subsection{Full model}

\begin{figure}[t]
\begin{center}
\includegraphics[width=0.25\linewidth]{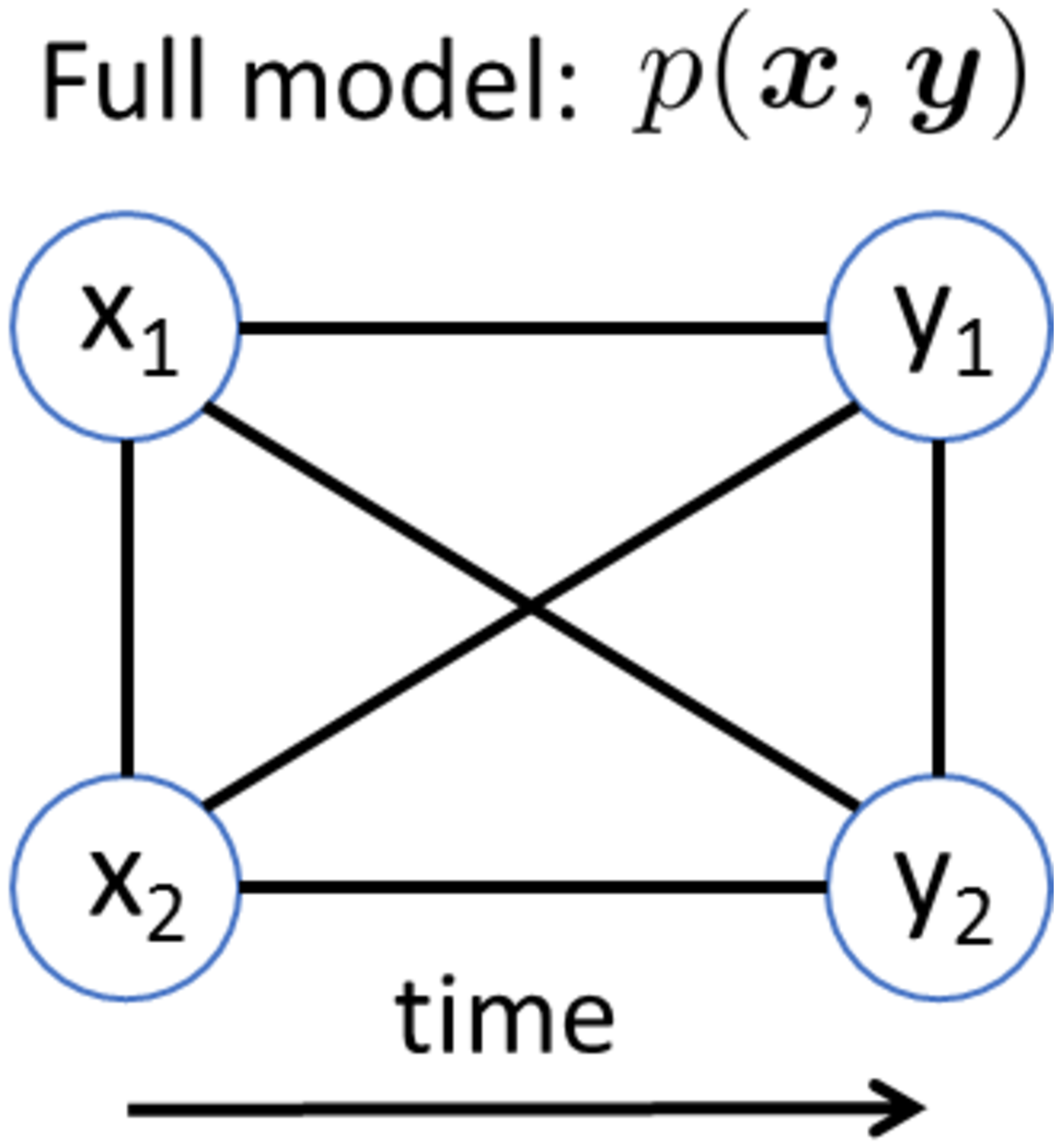} 
  \caption{Full model $p({\bm{x}}, {\bm{y}})$. }
  \label{fig:full_model}
\end{center}
\end{figure}

A full model $\mathcal{M}$, $p({\bm{x}}, {\bm{y}})$, is a graphical model in which all the nodes (terminals) are connected (Fig. \ref{fig:full_model}). We consider the binary case. In that case, $p({\bm{x}}, {\bm{y}})$ is an exponential family and can be expanded as
\begin{eqnarray}
 && p({\bm{x}}, {\bm{y}}) = \exp \biggl\{
 \sum \theta^X_i x_i + \sum \theta^Y_j y_j 
  + \sum \theta^{XX}_{ij} x_i  x_j \nonumber \\
  &&\mbox{\qquad\qquad\qquad\qquad}
  + \sum \theta^{YY}_{ij} y_i y_j + \sum \theta^{XY}_{ij} x_i y_j
  + h({\bm{x}}, {\bm{y}}) -\psi \biggr\},
\end{eqnarray}
where we show linear and quadratic terms explicitly by using parameters $\theta^X_i, \theta^Y_j, \theta^{XX}_{ij}, \theta^{YY}_{ij}, \theta^{XY}_{ij}$. $h({\bm{x}}, {\bm{y}})$ is the higher order terms of ${\bm{x}}$ and ${\bm{y}}$ and the last term $\psi$ is the free energy term (or cumulant generating function) corresponding to the normalizing factor. The set of distributions in the full model form a dually flat statistical manifold \cite{Amari2016}.

We hereafter neglect higher-order terms, since they disappear in split models we consider. Then, parameters
\begin{equation}
 {\bm{\theta}} = \left( \theta^X_i, \theta^Y_j, \theta^{XX}_{ij}, 
 \theta^{YY}_{ij}, \theta^{XY}_{ij} \right)
\end{equation}
form an $e$-coordinate system to specify a distribution $p({\bm{x}}, {\bm{y}})$. The dual coordinate system, $m$-coordinate system, is denoted by ${\bm{\eta}}$,
\begin{equation}
 {\bm{\eta}} = \left( \eta^X_i, \eta^Y_j, \eta^{XX}_{ij},
  \eta^{YY}_{ij}, \eta^{XY}_{ij} \right).
\end{equation}
The components of ${\bm{\eta}}$ are expectations of corresponding random variables. For example,
\begin{eqnarray}
 \eta^{XX}_{ij} &=& {\rm{E}} \left[ x_i x_j \right], \\
 \eta^{XY}_{ij} &=& {\rm{E}} \left[ x_i y_j \right],
\end{eqnarray}
where ${\rm{E}}$ is the expectation. In the followings, we consider the case where the number of elements is $2$ ($n=2$) for the explanatory purpose, but generalization for larger $n$ is straightforward.

\subsection{Fully split model}

\begin{figure}[t]
\begin{center}
\includegraphics[width=0.25\linewidth]{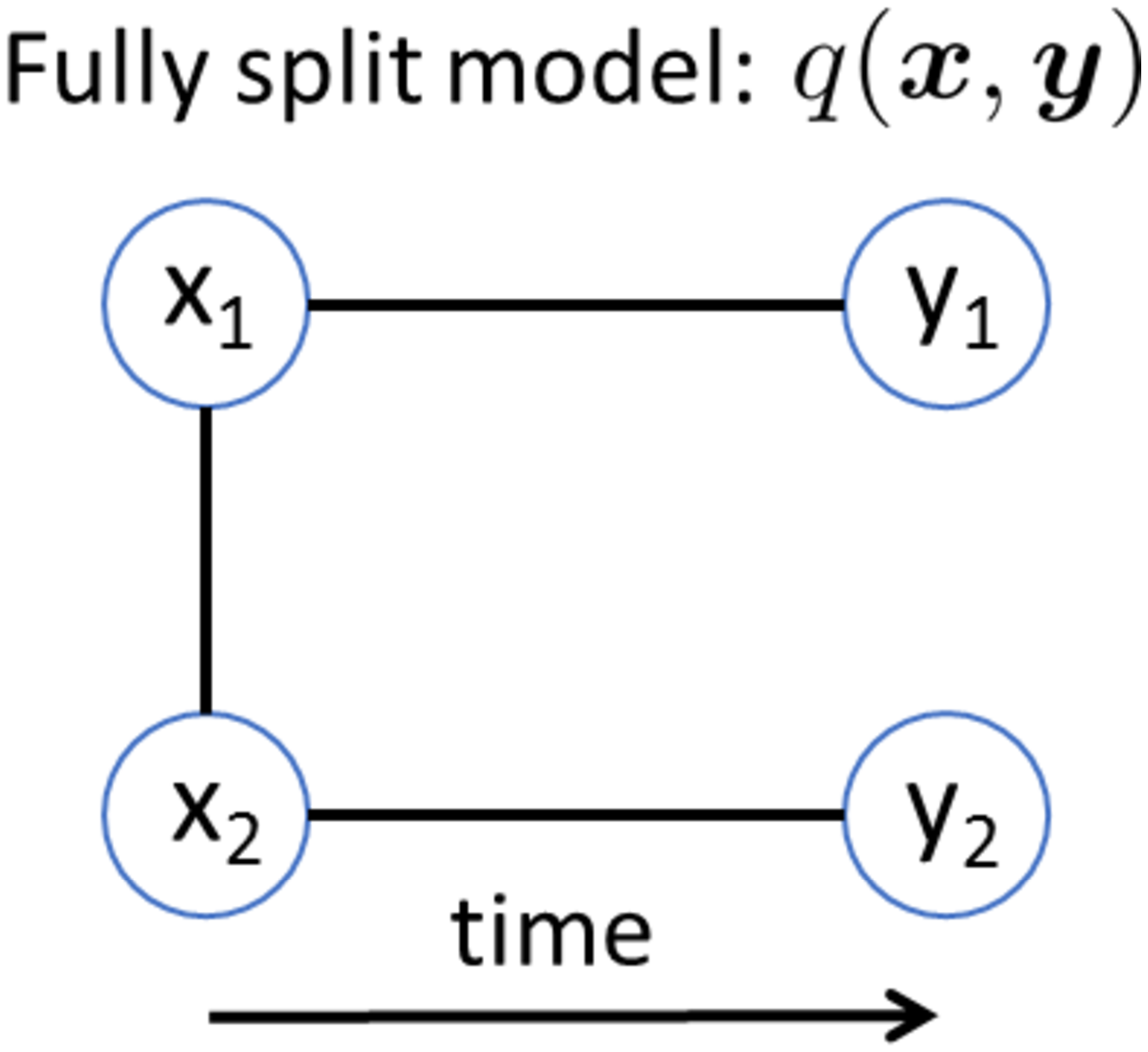} 
  \caption{Fully split model $q({\bm{x}},{\bm{y}})$.}
  \label{fig:fully_split_model}
\end{center}
\end{figure}

Ay considered a split model from the viewpoint of complexity of a system\cite{Ay2001,Ay2015}. The split model $q({\bm{y}}|{\bm{x}})$ is given by 
\begin{equation}
q({\bm{y}}|{\bm{x}}) = \prod_i q(y_i|x_i),
\end{equation}
where the conditional probability distribution of the whole system $q({\bm{y}}|{\bm{x}})$ is fully split into that of each part. We call this model ``fully split model'' $\mathcal{M}_{FS}$. The corresponding measure was also introduced by Barrett and Seth \cite{Barrett2010} following the measure of integrated information proposed by Balduzzi and Tononi \cite{Balduzzi2008}. 

This split model deletes branches connecting $X_i$ and $Y_j$ ($i \ne j$) and also deletes the branches connecting different $Y_i$ and $Y_j$ ($i \ne j$) (Here, we use capital letters $X$ and $Y$ to emphasize random variables, not their values.). This split model is reasonable because when terminals $Y_i$ are split, all the branches connecting $Y_i$ and the other nodes should be deleted except for branches connecting $X_i$ and $Y_i$. Branches connecting $X_i$ and $X_j$ remain as they are (Fig. \ref{fig:fully_split_model}). However, even though branches connecting $Y_i$ and $Y_j$ are deleted, this does not imply that $Y_i$ and $Y_j$ ($i \ne j$) are independent, because when input $X_i$ and $X_j$ are correlated, $Y_i$ and $Y_j$ are also correlated even though no branches exist connecting $X_i$ and $Y_j$ and $Y_i$ and $Y_j$. Even if branches connecting $Y_i$ and $Y_j$ are deleted, however, it does not imply that $Y_i$ and $Y_j$ ($i \ne j$) are independent; when input $X_i$ and $X_j$ are correlated, $Y_i$ and $Y_j$ are also correlated without any branches connecting $X_i$ and $Y_j$ and $Y_i$ and $Y_j$.

When $n=2$, the random variables $X_i$ and $Y_j$ have a Markovian structure,
\begin{equation}
 Y_1-X_1-X_2-Y_2,
\end{equation}
so that $Y_1$ and $Y_2$ are conditionally independent when $\left(X_1,X_2\right)$ is fixed. Also $X_2$ and $Y_1$ (or $X_1 \;\mbox{and}\; Y_2$) are conditionally independent when $X_1$ (or $X_2$) are fixed. These constraints correspond to putting
\begin{equation}
 \label{eq:am13-20170302}
 \theta^{XY}_{12} = \theta^{XY}_{21} = \theta^{YY}_{12} = 0
\end{equation}
in the ${\bm{\theta}}$-coordinates. They are linear constraints in the ${\bm{\theta}}$-coordinates. Thus, the fully split model $\mathcal{M}_{FS}$ is an exponential family. It is an $e$-flat submanifold of $\mathcal{M}$. Given $p({\bm{x}}, {\bm{y}}) \in \mathcal{M}$, let $q^{\ast}({\bm{x}}, {\bm{y}})$ be the $m$-projection of $p$ to $\mathcal{M}_{FS}$. Then, $q^{\ast}({\bm{x}}, {\bm{y}})$ is given by the minimizer of KL-divergence,
\begin{equation}
 q^{\ast}({\bm{x}}, {\bm{y}}) = \argmin_{q({\bm{x}}, {\bm{y}}) \in M_{FS}} D_{KL}[p({\bm{x}}, {\bm{y}}): q({\bm{x}}, {\bm{y}})].
\end{equation}
We use the mixed coordinate system of $\mathcal{M}$,
\begin{equation}
 {\bm{\xi}} = \left( \eta^X_i, \eta^Y_j, \eta^{XX}_{ij}, \eta^{YY}_{ij},
  \eta^{XY}_{11}, \eta^{XY}_{22} \;;\; \theta^{XY}_{12}, \theta^{XY}_{21},
  \theta^{YY}_{12} \right).
\end{equation}
Then $\mathcal{M}_{FS}$ is specified by (\ref{eq:am13-20170302}).

Because of the Pythagorean theorem, the $m$-projection of $p$ to $\mathcal{M}_{FS}$ that minimizes the $KL$-divergence $D_{KL} \left[p : \mathcal{M}_{FS} \right]$ is explicitly given by
\begin{equation}
 {\bm{\xi}}^{\ast} = \left(
  \eta^X_i, \eta^Y_j, \eta^{XX}_{ij}, \eta^{YY}_{ij},
  \eta^{XY}_{11}, \eta^{XY}_{22} \;;\; 0 \right)
\end{equation}
in the ${\bm{\xi}}$-coordinate system, where ${\bm{\eta}}$-part is the same as that of the mixed coordinates of $p({\bm{x}}, {\bm{y}})$.

By simple calculations, we obtain
\begin{equation}
 q^{\ast}({\bm{x}}, {\bm{y}})= p({\bm{x}})p
 \left(y_1|x_1\right) p \left(y_2|x_2 \right),
\end{equation}
which means
\begin{eqnarray}
 q^{\ast}({\bm{x}}) &=& p({\bm{x}}), \\
 q^{\ast}({\bm{y}}|{\bm{x}}) &=& \prod p \left(y_i |x_i \right).
\end{eqnarray}
The corresponding measure of integrated information is given by
\begin{equation}
 \Phi_{FS} = \sum H \left[Y_i | X_i \right] - H[Y|X], \label{eq:SI}
\end{equation}
where $H \left[Y_i|X_i \right]$ and $H[Y|X]$ are the conditional entropies corresponding to the random variables. This measure was termed ``stochastic interaction'' by Ay \cite{Ay2001}.

While $\Phi_{FS}$ is straightforward in derivation and its concept, it has an undesirable property as a measure of integrated information. Specifically, as we proposed in \cite{Oizumi2016PLoS,Oizumi2016PNAS}, any measure of integrated information $\Phi$, is expected to satisfy the following constraint,
\begin{equation}
 \label{eq:am21-20170302}
 0 \le \Phi \le I (X;Y),
\end{equation}
where $I(X;Y)$ is the mutual information between $X$ and $Y$. This requirement is natural because $\Phi$ should quantify the ``loss of information'' caused by splitting a system into parts, i.e., removing information transmission between parts. The loss of information should not exceed the total amount of information in the whole system, $I(X;Y)$, and should be always positive or 0. $\Phi$ should be 0 only when $X$ and $Y$ are independent. However, $\Phi_{FS}$ does not satisfy the requirement of the upper bound, as was pointed by \cite{Oizumi2016PLoS,Oizumi2016PNAS}. This is because $\mathcal{M}_{FS}$ does not include the submanifold $\mathcal{M}_I$ consisting of the independent distributions of $X$ and $Y$,
\begin{equation}
 \mathcal{M}_I = \left\{ q({\bm{x}}) q({\bm{y}})\right\}. \label{eq:M_I}
\end{equation}
$M_I$ is characterized by
\begin{equation}
 \theta^{XY}_{ij} = 0 \quad \textrm{(for $\forall$ $i$,$j$)}. 
\end{equation}
It is an $e$-flat submanifold of $\mathcal{M}$. The minimized KL-divergence between $p({\bm{x}},{\bm{y}})$ and $\mathcal{M}_I$ is mutual information,
\begin{equation}
I(X;Y) = \min_{q \in \mathcal{M}_I} D_{KL}(p({\bm{x}},{\bm{y}}):q({\bm{x}},{\bm{y}}))
\end{equation}

Thus, while stochastic interaction, derived from the submanifold $\mathcal{M}_{FS}$, has a simple expression (Eq. \ref{eq:SI}) and nice properties on its own, it may not be an ideal measure of integrated information due to its violation of the upper-bound requirement.

\subsection{Diagonally split graphical model}

\begin{figure}[t]
\begin{center}
  \includegraphics[width=0.25\linewidth]{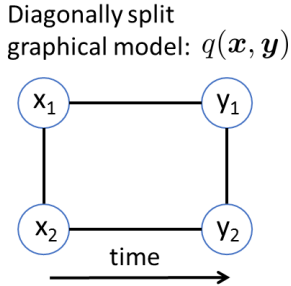} 
  \caption{Diagonally split graphical model $q({\bm{x}}, {\bm{y}})$. }
  \label{fig:diagonally_split_model}
\end{center}
\end{figure}

In order to overcome the above difficulties, we consider an undirected graphical model in which all the branches connecting $x_i$ and $y_j (i \ne j)$ are deleted but all the other branches remain as shown in Fig. \ref{fig:diagonally_split_model}. We call this model ``diagonally split graphical model'' $\mathcal{M}_{DS}$. 

The model is defined by
\begin{equation}
 \theta^{XY}_{12} = \theta^{XY}_{21} = 0,
\end{equation}
It is also an $e$-flat submanifold of $\mathcal{M}$. The branches connecting different $y_i$ exist so that $\theta^{YY}_{ij} \ne 0$. The model does not remove direct interactions among $y_i$, which can be caused by correlated noises directly applied to the output nodes (not through causal influences from $\bm{x}$). The fully split model $\mathcal{M}_{FS}$ introduced in the previous section is an $e$-flat submanifold of $\mathcal{M}_{DS}$, since $\theta^{YY}_{ij}=0 \; (i \ne j)$ is further required for $M_{FS}$.

In the case of $n=2$, the full model $\mathcal{M}$ is 10-dimensional (excluding higher-order interactions), $\mathcal{M}_{FS}$ is 7-dimensional and $\mathcal{M}_{DS}$ is 8-dimensional. $\mathcal{M}_{DS}$ satisfies the conditions that ${\bm{x}}_1$ and ${\bm{y}}_2$ as well as ${\bm{x}}_2$ and ${\bm{y}}_1$ are conditionally independent when $\left({\bm{x}}_2, {\bm{y}}_1\right)$ and $\left({\bm{x}}_1, {\bm{y}}_2\right)$ are fixed, respectively. However, no Markovian type relations hold because the graph is cyclic. The model is characterized by
\begin{equation}
 q({\bm{x}}, {\bm{y}}) = f({\bm{x}}) g({\bm{y}}) \prod_i h \left(x_i, y_i\right).
\end{equation}

We use the following mixed coordinates
\begin{equation}
 {\bm{\xi}} = \left(
  \eta^X_i, \eta^Y_j, \eta^{XX}_{ij}, \eta^{YY}_{ij} \;;\;
  \theta^{XY}_{12}, \theta^{XY}_{21}
 \right).
\end{equation}
Then, the $m$-projection of $p({\bm{x}},{\bm{y}})$ to $\mathcal{M}_{SG}$ is given by
\begin{equation}
 {\bm{\xi}}^{\ast} = \left(
  \eta^X_i, \eta^Y_j, \eta^{XX}_{ij}, \eta^{YY}_{ij} \;;\; 0,0
 \right)
\end{equation}
in these coordinates. This implies that
\begin{eqnarray}
 q^{\ast}({\bm{x}}) &=& p({\bm{x}}), \\
 q^{\ast}({\bm{y}}) &=& p({\bm{y}}), \\
 q^{\ast} \left(y_i | x_i \right) &=& p \left(y_i | x_i \right),
 \quad \forall i.
\end{eqnarray}
The corresponding measure of integrated information is
\begin{equation}
 \Phi_{DS} = D_{KL} \left[ p({\bm{x}}, {\bm{y}}):
 q^{\ast}({\bm{x}}, {\bm{y}})\right]. \label{eq:Phi_DS}
\end{equation}
It satisfies the natural requirement for integrated information (Eq. \ref{eq:am21-20170302}). Thus, it resolves the shortcomings of $\Phi_{FS}$.

However, there still remains a problem to take into consideration. To illustrate it, let us consider the two terminal Gaussian case (autoregressive (AR) model), in which ${\bm{x}}$ is linearly transformed to ${\bm{y}}$ by the connectivity matrix $A$ and the Gaussian noise $\bm{\epsilon}$ is added,
\begin{equation}
 {\bm{y}} = A{\bm{x}} + {\bm{\epsilon}}.
\end{equation}
Here, in the two terminals case, $A$ is given by,
\begin{equation}
 A = \left[
  \begin{array}{cc}
   A_{11} & A_{12} \\
   A_{21} & A_{22}
  \end{array}
 \right],
\end{equation}
and $\bm{\epsilon}$ is zero mean Gaussian noise whose covariance matrix is given by
\begin{equation}
 \Sigma(E) = \left[
  \begin{array}{cc}
   \sigma^2_{1} & \sigma_{12} \\
   \sigma_{21} & \sigma^2_2
  \end{array}
 \right].
\end{equation}
Let $\Sigma(X)$ be the covariance matrix of ${\bm{x}}$.  Then, the joint
probability distribution is written as
\begin{equation}
 \label{eq:am35-20170302}
 p({\bm{x}}, {\bm{y}}) = \exp \left\{
  -\frac 12 \left({\bm{x}}^T \Sigma(X)^{-1} {\bm{x}} \right) + 
  ({\bm{y}}-A{\bm{x}})^T \Sigma(E)^{-1} ({\bm{y}}-A{\bm{x}}) -\psi 
 \right\},
\end{equation}
where the means of all random variables are assumed to be equal to 0. The ${\bm{\theta}}$-coordinates consist of three matrices,
\begin{eqnarray}
 \label{eq:am36-20170302}
 {\bm{\theta}} &=& \left( \theta_{XX}, \theta_{YY}, \theta_{XY}\right), \\
 \label{eq:am37-20170302}
 \theta_{XX} &=& \Sigma(X)^{-1}, \quad \theta_{YY} = \Sigma(E)^{-1},
 \\
 \label{eq:am38-20170302}
 \theta_{XY} &=& -A \Sigma(E)^{-1}
\end{eqnarray}
and the corresponding ${\bm{\eta}}$-coordinates are
\begin{eqnarray}
 {\bm{\eta}} &=& \left( \eta_{XX}, \eta_{YY}, \eta_{XY} \right), \\
 \eta_{XX} &=& \Sigma(X), \quad
 \eta_{YY} = A \Sigma(X) A, \quad
 \eta_{XY} = A \Sigma(X).
\end{eqnarray}

We project $p({\bm{x}}, {\bm{y}})$ (Eq. \ref{eq:am35-20170302}) to $\mathcal{M}_{DS}$. The closest point $q^{\ast}({\bm{x}}, {\bm{y}})$ is again given by an AR model,
\begin{equation}
 {\bm{y}} = A^{\ast}{\bm{x}} + {\bm{\varepsilon}}^{\ast}.
\end{equation}
where $A^{\ast}$ and the covariance matrix of ${\bm{\varepsilon}}^{\ast}$, $\Sigma(E^{\ast})$, are determined from $A$, $\Sigma(X)$ and $\Sigma(E)$. However, the off-diagonal elements of $A^{\ast}$ is not zero. Therefore, the deletion of the diagonal branches in a graphical model is not equivalent to the deletion of the off-diagonal elements of $A$ in the Gaussian case. 
%
%
%

The off-diagonal elements of $A$, $A_{ij}$, determines causal influences from $x_i$ to $y_j$. In the diagonal split model $\mathcal{M}_{DS}$, the causal influences are non-zero because the off-diagonal elements of $A^*$, $A^*_{ij}$, are non-zero. Thus, the corresponding measure of integrated information $\Phi_{DS}$ (Eq. \ref{eq:Phi_DS}) does not purely quantify causal influences between the elements. In IIT, integrated information is designed to quantify causal influences \cite{Balduzzi2008,Oizumi2014}. In this sense, it is desirable to have a split model, which results in a diagonal connectivity matrix $A$. 

\subsection{Causally split model (Geometric model)}

To derive a split model where only causal influences between elements are removed, we consider that the essential part is to remove branches connecting $x_i$ and $y_j \; (i \ne j)$, without destroying other constituents. The minimal requirement to remove the effect of the branch $(i, j)$ is to let $x_i$ and $y_j$ be conditionally independent, when all the other elements are fixed. In our case of $n=2$, we should have two Markovian conditions
\begin{eqnarray}
 \label{eq:am45-20170302}
  X_1 \mbox{---} X_2 \mbox{---} Y_2, \\
 \label{eq:am46-20170302}
  X_2 \mbox{---} X_1 \mbox{---} Y_1.
\end{eqnarray}
The split model that satisfies the above conditions was introduced by Oizumi, Tsuchiya and Amari \cite{Oizumi2016PNAS} and was called ``geometric model'' $\mathcal{M}_G$, because information geometry was used as a guiding principle to obtain the model. We can also call it ``causally split model'' because causal influences between elements are removed.

The model $\mathcal{M}_G$ is a 8-dimensional submanifold of $\mathcal{M}$ in the case of $n=2$, because there are two constraints (Eqs. \ref{eq:am45-20170302} and \ref{eq:am46-20170302}). These constraints are expressed as
\begin{eqnarray}
 q \left(x_1, y_2 | x_2 \right) &=& 
 q \left(x_1|x_2\right) q \left(y_2|x_2\right), \\
 q \left(x_2, y_1 | x_1 \right) &=&
 q \left(x_2 | x_1\right)q \left(y_1|x_1\right).
\end{eqnarray}
We can write down the constraints in terms of ${\bm{\theta}}$-coordinates, but they are nonlinear. They are also nonlinear in the ${\bm{\eta}}$-coordinates. Thus, $\mathcal{M}_G$ is a curved submanifold and it is not easy to give an explicit solution of the $m$-projection of $p({\bm{x}}, {\bm{y}})$ to $\mathcal{M}_G$.

%
%

We can solve the Gaussian case explicitly \cite{Oizumi2016PNAS}. It is not difficult to prove that, when the Markovian conditions in Eqs. \ref{eq:am45-20170302} and \ref{eq:am46-20170302} are satisfied, the connectivity matrix $A'$ of an AR model in $\mathcal{M}_G$,
\begin{equation}
 {\bm{y}} = A' {\bm{x}} + E',
\end{equation}
is a diagonal matrix. From (\ref{eq:am38-20170302}), we have
\begin{equation}
 A' = -\theta^{-1}_{YY} \theta_{XY}.
\end{equation}
Thus, the constraints in Eqs. \ref{eq:am45-20170302} and \ref{eq:am46-20170302} expressed in terms of ${\bm{\theta}}$-coordinates are equivalent to the off-diagonal elements of matrix
$\theta^{-1}_{YY} \theta_{XY}$ being 0. Thus, the constraints are nonlinear in the ${\bm{\theta}}$-coordinates. The corresponding measure of integrated information, $\Phi_G$ (geometric integrated information), is given explicitly by
\begin{equation}
 \Phi_G = \frac{1}{2} \log \frac{\left|
 \Sigma(E) \right|}{ \left|\Sigma(E') \right|},
\end{equation}
where $\Sigma(E)$ is the noise covariance of $p({\bm{x}}, {\bm{y}})$, $\Sigma(E')$ is that of projected $q^{\ast}({\bm{x}}, {\bm{y}})$, $|\Sigma(E)|$ is the determinant of $\Sigma(E)$. 

By construction, it is easy to see that $\Phi_G$ satisfies the requirements for integrated information,
\begin{equation}
 0 \le \Phi_G \le I(X; Y),
\end{equation}
because the causally split model $\mathcal{M}_G$ includes the submanifold $\mathcal{M}_I$ consisting of the independent distributions of $X$ and $Y$ (Eq. \ref{eq:M_I}). We believe that $\Phi_G$ is the best candidate measure in the sense that it is closest to the original philosophy of integrated information in IIT. 
In IIT, integrated information is designed to quantify causal influences between elements \cite{Balduzzi2008,Oizumi2014}. Note that in IIT, ``causal'' influences are quantified by Pearl's intervention framework \cite{Pearl2009, Balduzzi2008, Ay2008} attempting to quantify the ``actual'' causation. On the other hand, causal influences quantified in this paper do not necessarily mean actual causation. $\Phi_G$ is related to observational measures of causation such as Granger causality or Transfer entropy \cite{Oizumi2016PNAS}.

%
%

\subsection{Mismatched decoding model}

As a different direction from the above measures of integrated information, we can consider another model, called a mismatched decoding model $\mathcal{M}_{MD}$. We use the concept of mismatched decoding in information theory proposed by Merhav et al \cite{Merhav1994}. We have utilized this concept in the context of neuroscience \cite{Oizumi2010,Oizumi2011,Boly2015, Oizumi2016PLoS,Haun2017}. 

To introduce the decoding perspective, let us consider a situation where we try to estimate the input ${\bm{x}}$ when the output ${\bm{y}}$ is observed. When we know the correct joint probability distribution $p({\bm{x}}, {\bm{y}})$, we can estimate ${\bm{x}}$ by using the true distribution $p({\bm{x}}|{\bm{y}})$. This is the optimal matched decoding. However, when we use a split model $q({\bm{x}}, {\bm{y}})$ for decoding, there is always loss of information. This type of decoding is called mismatched decoding because the decoding model $q({\bm{x}}, {\bm{y}})$ is different from the actual probability distribution $p({\bm{x}}, {\bm{y}})$. 

We previously considered the fully split model as a mismatched decoding model \cite{Oizumi2016PLoS}
\begin{equation}
q({\bm{y}}|{\bm{x}}) = \prod_i q(y_i|x_i).
\end{equation}
By using the Merhav's framework, the information loss when $q({\bm{y}}|{\bm{x}})$ is used for decoding can be quantified by \cite{Merhav1994,Oizumi2016PLoS}
\begin{equation}
 \Phi_{MD} = {\mathop{\min}_{\beta}} D_{KL} 
 \left[ p({\bm{x}}, {\bm{y}}) || q({\bm{x}}, {\bm{y}} ; \beta)\right].
\end{equation}
where
\begin{equation}
 q({\bm{x}}, {\bm{y}}; \beta) = \frac{p({\bm{x}}) p({\bm{y}}) \prod_i p \left( y_i | x_i \right)^{\beta}}{\sum_{\bm{x}'} p({\bm{x'}}) \prod_i p \left( y_i | x'_i \right)^{\beta}}. \label{eq:M_MD}
\end{equation}
To quantify the information loss $\Phi_{MD}$, the KL-divergence needs to be minimized with respect to the one-dimensional parameter $\beta$. We call $q({\bm{x}}, {\bm{y}}; \beta)$ ``mismatched decoding model'' $\mathcal{M}_{MD}$. The mismatched decoding model $\mathcal{M}_{MD}$ forms one-dimensional submanifold. As can be seen in Eq. \ref{eq:M_MD}, no interaction terms are included between $x_i$ and $y_j \; (i \ne j)$. Thus, $\mathcal{M}_{DM}$ is included in the diagonally split graphical model $\mathcal{M}_{DS}$. 

The optimal $\beta^{\ast}$, which minimizes the KL-divergence, is given by projecting $p({\bm{x}}, {\bm{y}})$ to $\mathcal{M}_{MD}$. Since $\mathcal{M}_{MD}$ is not an $e$-flat submanifold, it is difficult to obtain the analytical expression of $q^{\ast}({\bm{x}}, {\bm{y}})$. However, the minimization of KL-divergence is a convex problem and thus, the optimal $\beta^*$ can be easily found by numerical calculations such as gradient descent \cite{Latham2005,Oizumi2016PLoS}.

\section{Comparison of various measures of integrated information}

We have derived four  measures of integrated information from four different definitions of the split model. We elucidate their relations in this section.  

First, $\mathcal{M}_{FS}$ and $\mathcal{M}_{DS}$ are $e$-flat submanifolds, forming exponential families.  Therefore, we can directly apply the Pythagorean projection theorem and the projected $q^{\ast}({\bm{x}}, {\bm{y}})$ is explicitly obtained by using the mixed coordinates. However, $\mathcal{M}_G$ and $\mathcal{M}_{MD}$ are curved submanifolds and thus, it is difficult to analytically obtain the projected $q^{\ast} ({\bm{x}}, {\bm{y}})$ in general.

The natural requirements for integrated information, 
\begin{equation}
 0 \le \Phi \le I (X, Y),
\end{equation}
are satisfied for all the measures of integrated information except for $\mathcal{M}_{FS}$. This is because $\mathcal{M}_{FS}$ does not include $\mathcal{M}_{I}$ (Eq. \ref{eq:M_I}) while the other split models include $\mathcal{M}_{I}$. 

In general, when $\mathcal{M}_1 \supset \mathcal{M}_2$,
\begin{equation}
\min_{q \in \mathcal{M}_1} D_{KL} \left[ p:q \right] \le \min_{q \in \mathcal{M}_2} D_{KL} \left[ p: q \right]
\end{equation}
and therefore, 
\begin{equation}
 \Phi_2 \ge \Phi_1.
\end{equation}
We have proved 
\begin{eqnarray}
 && \mathcal{M}_{DS} \supset \mathcal{M}_{FS}, \;  \mathcal{M}_{DS} \supset \mathcal{M}_{MD}, \\
 && \mathcal{M}_G \supset \mathcal{M}_{FS}.
\end{eqnarray}
From these relations between the split models, we have the relations between the corresponding measures of integrated information,
\begin{equation}
 \Phi_{FS} \ge \Phi_{DS}, \quad
 \Phi_{MD} \ge \Phi_{DS}, \quad
 \Phi_{FS} \ge \Phi_G.
\end{equation}
$\mathcal{M}_{FS}$ is included in the intersection of $\mathcal{M}_{DS}$ and $\mathcal{M}_G$. $\mathcal{M}_I$ is included in $\mathcal{M}_{DS}$, $\mathcal{M}_G$, and $\mathcal{M}_{MD}$.

\begin{figure}[t]
\begin{center}
  \includegraphics[width=0.7\linewidth]{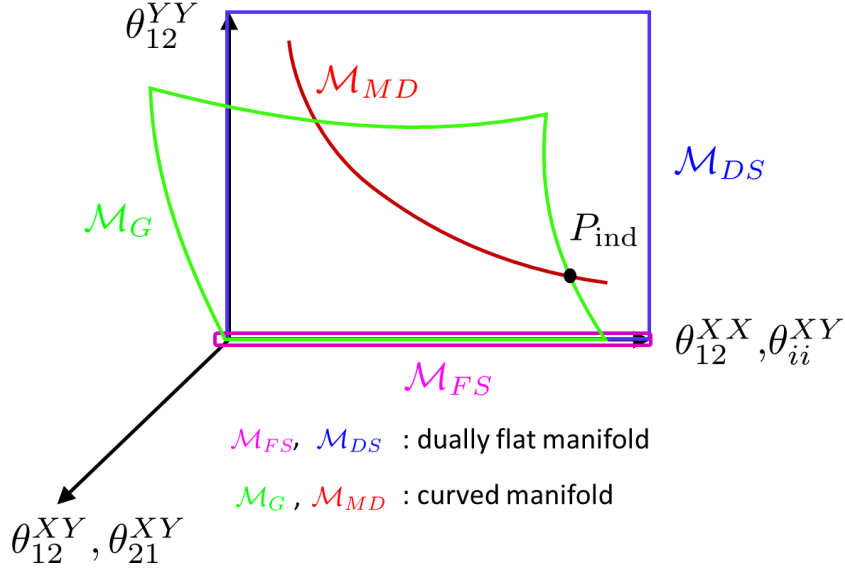} 
  \caption{Relations among four different split models. Fully split model $\mathcal{M}_{FS}$ and diagonally split graphical model $\mathcal{M}_{DS}$ are dually flat manifolds. $\mathcal{M}_{FS}$ is represented by a magenta line on the axis of $\theta_{12}^{XX}, \theta_{ii}^{XY}$. $\mathcal{M}_{DS}$ is represented by a blue square spanned by the two axes $\theta_{12}^{XX}, \theta_{ii}^{XY}$ and $\theta_{12}^{YY}$. Causally split model (geometric model) $\mathcal{M}_G$ and mismatched decoding model $\mathcal{M}_{MD}$ are curved manifolds. $\mathcal{M}_G$ is represented by a curved green surface. $\mathcal{M}_{MD}$ is represented by a curved red line inside the surface of $\mathcal{M}_{DS}$. $P_{\rm ind}$ is an independent distribution of $\bm{x}$ and $\bm{y}$, $P_{\rm ind}=p({\bm{x}})p({\bm{y}})$, which is represented by a black point. $P_{\rm ind}$ is included in $\mathcal{M}_{DS}$, $\mathcal{M}_{MD}$, and $\mathcal{M}_{G}$ but is not included in $\mathcal{M}_{FS}$. }
 \label{fig:phi_relations}
\end{center}
\end{figure}

The relations among four different measures of integrated information are schematically summarized in Fig. \ref{fig:phi_relations}.

%
%
%
%
%
%
%

\section{Conclusions}

We studied four different measures of integrated information in a causal stochastic dynamical system from the unified viewpoint of information geometry. The four measures have their own meanings and characteristics. We elucidated their relations and a hierarchical structure of the measures (Fig. \ref{fig:phi_relations}). We can define a measure of information transfer for each branch, but their effects are not additive but subadditive. Therefore, we need to study further collective behaviors of deleting branches \cite{Jost2007}. This remains a future problem to be studied.
%
%

\bibliographystyle{plos2015}

\end{document}